\documentclass{epl}

\title{Field-enhanced critical parameters in magnetically nanostructured
superconductors}

\author{M. V. Milo\v{s}evi\'{c} \and F. M. Peeters}

\institute{Departement Fysica, Universiteit Antwerpen (Campus Drie Eiken), \\
Universiteitsplein 1, B-2610 Antwerpen, Belgium}





\shorttitle{Field-enhanced critical parameters}

\begin{document}
\maketitle

\begin{abstract}
Within the phenomenological Ginzburg-Landau theory, we demonstrate
the {\it enhancement} of superconductivity in a superconducting
film, when nanostructured by a lattice of magnetic particles.
Arrays of out-of-plane magnetized dots (MDs) extend the critical
magnetic field and critical current the sample can sustain, due to
the interaction of the vortex-antivortex pairs and surrounding
supercurrents induced by the dots and the external flux lines.
Depending on the stability of the vortex-antivortex lattice, a
peak in the $H_{ext}-T$ boundary is found for applied integer and
rational matching fields, which agrees with recent experiments
[Lange {\it et al.} Phys. Rev. Lett. {\bf 90}, 197006 (2003)]. Due
to compensation of MDs- and $H_{ext}$-induced currents, we predict
the field-shifted $j_{c}-H_{ext}$ characteristics, as was actually
realized in previous experiment but not commented on [Morgan and
Ketterson, Phys. Rev. Lett. {\bf 80}, 3614 (1998)].
\end{abstract}

\section{Introduction}

Arrays of nanoscale ferromagnetic (FM) particles are potential
devices for applying well-defined local magnetic fields. When such
a nano-engineered magnetic lattice is combined with a
superconducting (SC) film, various new phenomena occur. Only
recently, submicrometer lithographic techniques have been
developed that allow to reduce the size of magnetic inclusions to
a scale comparable with the characteristic lengths of conventional
superconductors \cite{schuller}. Because of its technological
relevance, flux pinning in these FM/SC heterostructures has been
the subject of a vast amount of theoretical and experimental work
(see \cite{reichh,schuller2,vanbael} and references therein).
Introducing magnetic dot lattices has proven to be a very useful
tool to understand the plethora of physical effects related to the
interactions between vortices and material imperfections,
including matching effects or collective locking of the flux
lattice to the magnetic dot array, with consequently higher
critical current \cite{morgan}.

Over the last years, the interest shifted towards the more
fundamental properties of magnetically textured superconductors.
For example, though often neglected, the stray field of the
particles may strongly modulate the order parameter in the
underlying superconductor. Recently \cite{misko}, the appearance
of various vortex-antivortex states was predicted. Lange {\it et
al.} \cite{lange} demonstrated experimentally that if such sample
is exposed to an additional homogeneous field, the
superconductivity in some parts of the sample is enhanced, due to
field-compensation effects. Moreover, in certain parameter-range,
one can even restore superconductivity by {\it adding} magnetic
field. Besides FM/SC structures, only (EuSn)Mo$_6$S$_8$, organic
$\lambda$-(BETS)$_{2}$FeCl$_{4}$ materials and HoMo$_6$S$_8$ show
this unconventional behavior \cite{fis}.

\begin{figure}[b]
\onefigure[height=5cm]{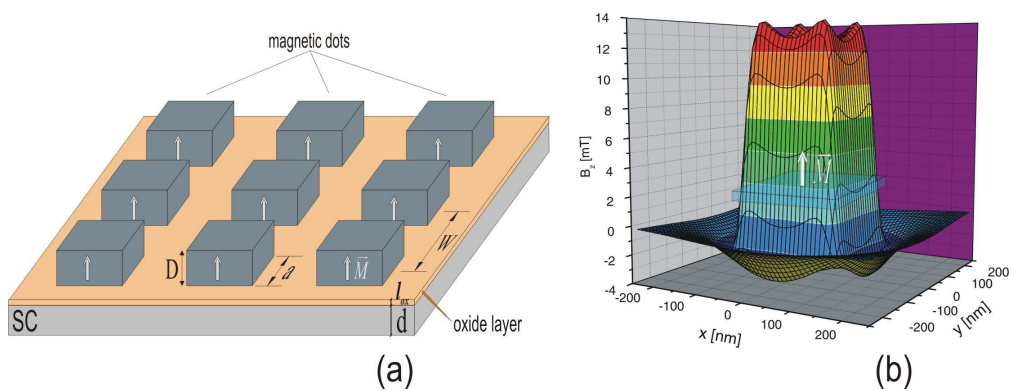}\caption{(a) The superconducting film
underneath a regular array of square magnetic dots with
out-of-plane magnetization. (b) The magnetic field profile under a
Co dot with $a=200$nm and $D=20$nm.} \label{fig:fig1}
\end{figure}

\section{Theoretical formalism}

The present Letter puts emphasis on the interplay of the magnetic
fields in a thin SC film with a square array of submicron magnetic
dots (MDs) with perpendicular magnetization (see Fig.
\ref{fig:fig1}), exposed to a background homogeneous field (in our
earlier works \cite{misko} no external magnetic field was
present). The superconductor and the magnetic array are only
magnetically coupled, as a thin oxide layer is assumed between
them to prevent the proximity effect. We investigate the influence
of the magnetization/stray field of the MDs on the critical
parameters of the superconductor. We present the first theoretical
(quantitative) explanation of the magnetic-field-induced
superconductivity. We broaden this physical picture and show how
the critical current in these samples can also be nanoengineered
by inducing additional currents in the sample by carefully
perpetrated MD-arrays.

In our theoretical treatment of this system, we use the non-linear
Ginzburg-Landau (GL) formalism, combined with specific boundary
conditions. The energy difference between the superconducting and
the normal state, in units of $H_{c}^{2}\big/4\pi$, can be
expressed as
\begin{eqnarray}
\Delta \mathcal{G}&=&\int \left[
-|\Psi|^{2}+\frac{1}{2}|\Psi|^{4}+\frac{1}{2}|(-i\nabla-{\bf A})
\Psi|^{2}+\kappa^{2}({\bf H}-{\bf H}_{0})^{2} \right]dV,
\label{freeen}
\end{eqnarray}
where ${\bf H}_{0}$ denotes the total magnetic field imposed on
the superconductor (magnetic dots plus external field).
Minimization of Eq. (\ref{freeen}) leads to two coupled GL
equations which we solve following a numerical approach proposed
by Schweigert {\it et al.} (see Ref. \cite{schweigert1}) on a
uniform Cartesian grid with typically 10 points/$\xi$ in each
direction. In the present case, we took for the simulation region
a rectangle $W_{x}\times W_{y}$, where $W_{x}=W_{y}=16W$ ($W$ is
the period of the MD lattice). In order to include periodicity of
SC and MD lattice in our calculation, we apply the periodic
boundary conditions \cite{doria}, with specific gauge
transformations given in Ref. \cite{misko}.

To explore the superconducting state, we start from randomly
generated initial configurations, increase/decrease the
magnetization of the MDs or change the value of the applied
external field, and let the vortex-configuration-solution relax to
a steady-state one. In addition, we always recalculate the vortex
structure starting from the Meissner state $(\Psi=1)$ or the
normal state $(\Psi\approx 0)$ as initial condition. By comparing
the energies of all found vortex states we determine the ground
state configuration. In this Letter, we are mostly interested in
the critical parameters for the superconducting/normal (S/N)
transition. In our calculations, the criterion
$|\psi|^{2}_{max}<10^{-5}$ denotes the normal state.

\section{The field-enhanced critical field}

\begin{figure}[b]
\onefigure[height=8cm]{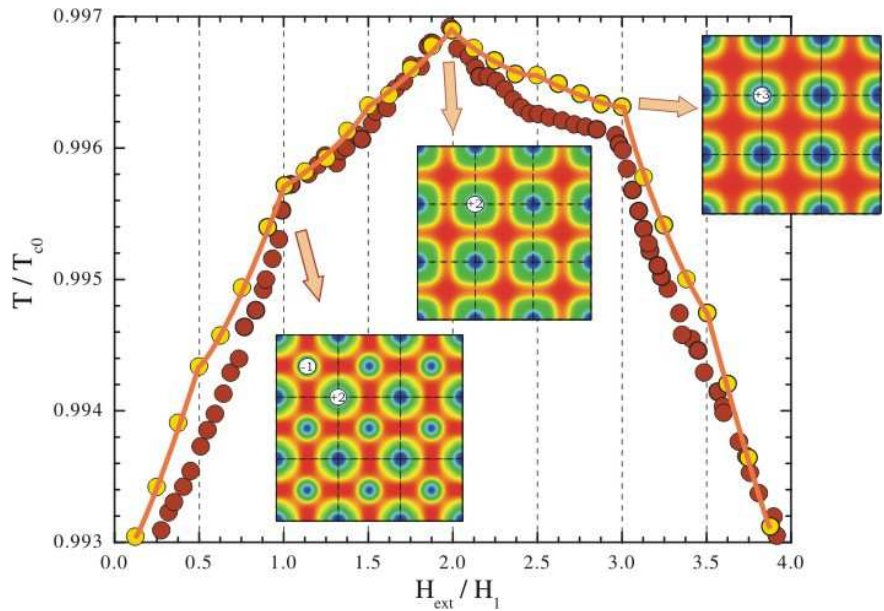}\caption{The $H_{ext}-T$ diagram: red
dots give the experimental data of Ref. \cite{lange}, and the
yellow ones correspond to the theoretical results. The Cooper-pair
density insets (blue/red color - low/high density) illustrate the
vortex configurations for the applied first and third matching
field (taken deeper inside the SC state). $T_{c0}$ denotes the
critical temperature in the absence of any magnetic field.}
\label{HTcomp}
\end{figure}
Recently, Lange {\it et al.} \cite{lange} demonstrated
experimentally the field-induced-superconductivity (FIS) effect. A
lattice of out-of-plane magnetic dots is placed on top of a SC
film (see Fig. \ref{fig:fig1}). The magnetic stray field of each
dot has a positive $z$-direction under the dots and a negative one
in the area between the dots. The basic idea is straightforward:
added to a homogeneous magnetic field $H_{ext}$, these dipole
fields {\it enhance} the $z$-component of the effective magnetic
field in the small area just under the dots and reduce the total
field everywhere else in the SC film (i.e. enhance
superconductivity). The sample consisted of a 85nm-thick Pb film,
covered by a 10nm Ge layer for protection from oxidation and
proximity effect. The square magnetic dots (side length about
0.8$\mu$m) were made as Pd(3.5nm)/[Co(0.4nm)/Pd(1.4nm)]$_{10}$
multilayers, and arranged in a square array with period 1.5$\mu$m.
Knowing values of these parameters, we can apply our numerical
approach to the investigated system. In Fig. \ref{HTcomp} the
calculated $H_{ext}-T$ diagram is shown for positively magnetized
dots. The temperature is introduced in our calculation through the
temperature dependence of the coherence length $\xi(T)=\xi(0)
\big/ \sqrt{1-T/T_{c}}$. The best agreement between the
experimental and theoretical $H_{ext}-T$ diagrams was obtained for
$\xi(0)\approx 28$nm and magnetization value of $M=3.32\times
10^{5}$A/m. While the magnetization corresponds to expected values
(between the Co and Pd values), the coherence length we found is
smaller than the known values for Pb films of 35-40nm. However,
$\xi$ is hardly a controllable quantity, and strongly depends on
the preparation of the sample. Notice that because of the
so-called ``virial theorem'' \cite{doria}, our approach works only
for integer number of external flux lines per simulation cell. In
this particular case, we used a $4\times 4$ supercell, so we were
able to calculate the phase boundary only in points described by
$H_{ext}=\frac{n}{16}H_{1}$, where $n$ is an integer number.

The $H_{ext}$-$T$ phase boundary is clearly altered by the
presence of the magnetic dot array. Contrary to a conventional
symmetric (with respect to $H_{ext}=0$) S/N phase boundary, the
$H_{ext}$-$T$ boundary for the magnetically textured SC is
strongly asymmetric, as shown in Fig. \ref{HTcomp}. The maximal
critical temperature was found for the second matching field
$H_{ext}=2H_{1}$ when $M>0$ ($H_{1}=9.2$G is the first matching
field, where number of vortices in the system matches the number
of magnetic pinning centers). The $H_{ext}$-$T$ phase diagram is
shifted over $\Delta H_{ext}=2H_{1}$ but at the same time is not
symmetric with respect to $H_{ext}=2H_{1}$. Due to this
applied-field-shift, the positive critical magnetic field of the
superconductor at given temperature is {\it enhanced} by magnetic
nanostructuring. However, note that the superconductivity becomes
less immune to the negative applied fields. If needed, that can be
accommodated by changing the polarity of the MDs, resulting in the
opposite field-shift effect.

Lange {\it et al.} estimated the negative flux (of the MD stray
field) between the magnetic dots to $\Phi^{-}\approx 2.1 \Phi_{0}$
per unit cell. From the field compensation effect described above,
one expects the maximal critical temperature when external flux
matches the flux of the negative stray field of the dots, which is
indeed the case in Fig. \ref{HTcomp}. However, this effect is not
related to the field-compensation, but to compensation of induced
currents in the sample, mostly through vortex-antivortex
annihilation. We found the same qualitative behavior for a range
of magnetization values of the dots, corresponding to the negative
flux range $\Phi^{-}/\Phi_{0}=2.21-3.22$ (maximal $T_{S/N}$ found
for $H_{ext}=2H_{1}$), while the best quantitative agreement with
experimental data was found for $\Phi^{-}=2.8\Phi_{0}$. In this
range of magnetization, each magnetic dot creates a $L=2$ giant
vortex under each dot and two antivortices at each interstitial
site \cite{misko}. Obviously, for the applied second matching
field, the external flux lines annihilate with the interstitial
antivortices, decreasing the total number of (anti)vortices in the
sample and giving rise to the critical temperature (see insets of
Fig. \ref{HTcomp}). Similar phenomenon happens for the first and
third matching field. For $H_{ext}=H_{1}$, one antivortex and one
external vortex annihilate, leaving one antivortex per unit cell
at the central interstitial position. This configuration is very
stable and leads to an increase of $T_{c}$. On the other hand, for
$H_{ext}=H_{3}$, the two antivortices annihilate with two external
vortices, and the third external flux line is pinned by the
magnetic dot, leading to a giant vortex with vorticity $L=3$ at
each pinning site. Therefore, temperature fluctuations affect this
vortex configuration much less than the one for $H_{ext}=H_{1}$
due to absence of weakly pinned interstitial (anti)vortices. As a
consequence, asymmetry in the $H_{ext}-T$ boundary with respect to
the second matching field is observed.
\begin{figure}[t]
\onefigure[height=5cm]{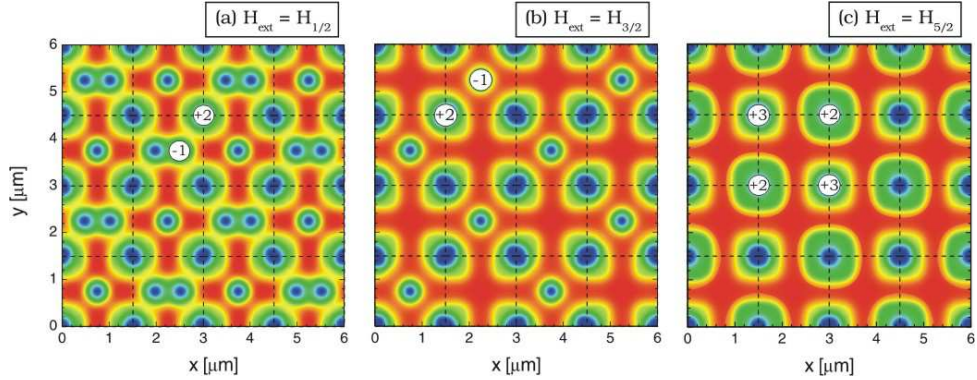} \caption{The Cooper-pair density
contourplots for the fractional vortex-antivortex lattices with
positive net vorticity, obtained for rational applied field: (a)
$H_{ext}=H_{1/2}$, (b) $H_{ext}=H_{3/2}$, and (c)
$H_{ext}=H_{5/2}$. Thin lines indicate the unit cells of the
magnetic dot lattice.} \label{frac}
\end{figure}

In addition, one should notice the fine structure in the $H-T$
boundary around the so-called rational applied magnetic fields.
Namely, we found small kinks in the phase boundary for external
fields $H_{i/2}$, where $i=1,~3,~5,~7$ (see Fig. \ref{HTcomp}).
For these fields, the number of external flux lines does not
correspond to an integer multiple of the number of antivortices
present in the sample. Therefore, after a very selective
annihilation, fractional vortex-antivortex configurations with
positive net vorticity per unit cell are formed. The Cooper-pair
density plots of such configurations are shown in Fig. \ref{frac}.
The newly formed vortex-antivortex lattices are still able to
preserve the symmetry, and due to the strong pinning of such a
lattice, an enhancement of superconductivity is observed. Note
that the unit cell of the vortex configurations for half-matching
fields is of size 2x2 lattice cells, which is different from the
matching field cases (see insets in Fig. \ref{HTcomp}). Note also
that each state shown in Fig. \ref{frac} exhibits certain
peculiarity: in (a) interstitial sites are alternately occupied by
1 or 2 antivortices; in (b) giant vortex ($L=2$) under each dot is
clearly deformed due to its attraction with antivortices at every
other interstitial site; in (c) there are no antivortices present,
and magnetic dots alternately capture 2 or 3 vortices in the form
of a giant vortex. However, the realization of fractional states
is rather difficult, as they are sensitive to the fluctuations in
the applied field (and other parameters determining competing
interactions). For that reason, theoretically found kinks in the
S/N boundary are hardly visible in the experimental data. However,
for well defined applied fields, we expect that these novel
vortex-antivortex states with non-zero total vorticity are
observable using e.g. scanning probe techniques like Hall and
Magnetic Force Microscopy.

\section{The field-shifted critical current characteristics}

\begin{figure}[t]
\onefigure[height=6cm]{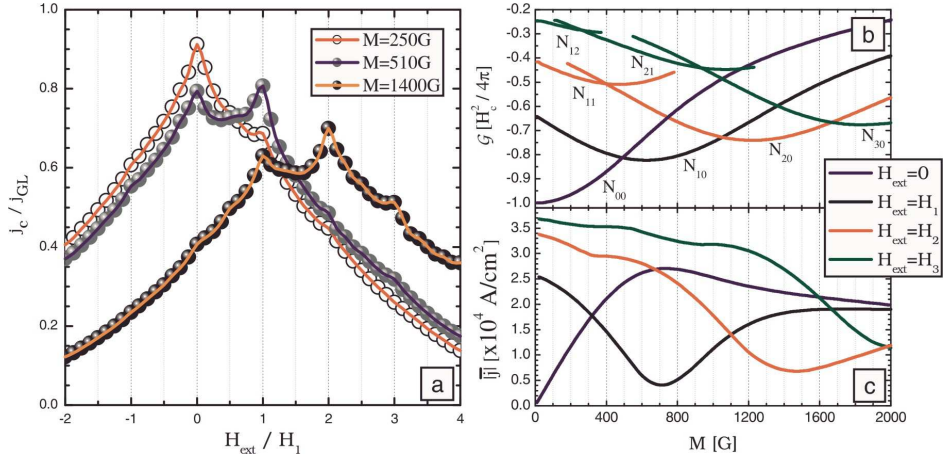} \caption{(a) Critical current (in
units of Ginzburg-Landau current $j_{GL}$) versus the applied
magnetic field in units of the first matching field. (b) The Gibbs
free energy and (c) the spatial average of the current magnitude
in the sample, as a function of the magnetization of the magnetic
dot array ($a=200$nm, $D=20$nm, $W=800$nm, above the Pb film at
$T/T_{c}=0.90$).} \label{fig:fig3}
\end{figure}

Magnetic nanostructuring of superconductors influences not only
their critical field but critical current as well. It is well
known that magnetic lattice on top of a SC film pins the external
flux lines when $M$ and $H_{ext}$ have the same polarity
\cite{vanbael,morgan}. According to Refs. \cite{vanbael,morgan},
this results in a reduced vortex mobility and consequently
enhanced critical current, with the maximal current for
$H_{ext}=0$. While this behavior is found for {\it weak} magnetic
pinning centers, more complicated physical picture is expected for
stronger ones. To investigate this, we exposed our (Pb) sample
with $a=200$nm, $D=20$nm, $W=800$nm and given magnetization $M$
and temperature $T/T_{c}=0.9$ to a gradually changed homogeneous
magnetic field, starting each time from the normal state. Then the
current is applied in the $x$-direction through $A_{cx}=const.$
(now $A_{0}=A_{md}+A_{ext}+A_{c}$) which does not interfere with
our boundary conditions, and the resulting current in the system
is calculated. When the critical value of $A_{cx}$ is reached, the
motion of (anti)vortices can no longer be prevented and
superconductivity is destroyed. The results of our calculations
for the critical current $j_{c}$ as a function of the applied
field are shown in Fig. \ref{fig:fig3}(a) for different values of
the MD-magnetization. For small magnetization (open dots) an
asymmetry in the $j_{c}(H_{ext})$ dependence with respect to
$H_{ext}=0$ is observed. This confirms the findings of Refs.
\cite{vanbael,morgan}, and indicates vortex pinning for parallel
orientation of the applied field and the magnetic moments. For
negative applied field, antivortices are introduced in the system
which are repelled by the magnetic dots. They are weakly pinned at
interstitial sites, further suppressing superconductivity and
easily stimulated to motion, leading to lower $j_{c}$.

However, for higher $M$ we found a field-offset in the
$j_{c}(H_{ext})$ characteristics. For $M=510$G and $M=1400$G (bulk
Ni and Co values), the maximal $j_{c}$ peak was found for
$H_{ext}=H_{1}$ and $H_{ext}=H_{2}$, respectively. Intuitively,
and according to our findings in previous section, we know that if
we increase $M$, vortex-antivortex (VAV) pairs nucleate. When
positive homogeneous field is applied, external flux lines
annihilate with the interstitial antivortices, and a maximal
critical current is obtained for a magnetic field such that all
antivortices are annihilated. In such a case, the field-offset in
the $j_{c}(H_{ext})$ curve indicates the number of VAV-pairs per
magnet. However, this mechanism is not always realized. The free
energy diagram in Fig. \ref{fig:fig3}(b) (blue curve) shows that
there are no vortex-antivortex pairs induced in the sample, even
for high $M$. Namely, vortices and antivortices cannot be
adequately separated since the magnetic lattice is too dense
compared to $\xi$(T) \cite{misko}. Nevertheless, when external
field is applied, the equilibrium vortex configurations (denoted
by $N_{ab}$, with a-number of pinned vortices per dot, b-number of
interstitial vortices) can have {\it lower} energy than the
superconducting state in the absence of the applied field. This
phenomenon occurs due to the current compensation effect, between
the magnet-induced antivortex-like currents and the currents of
external vortices, when pinned by the magnets. This feature is
obvious from Fig. \ref{fig:fig3}(c), where average magnitude of
current in the sample is shown as function of the magnet strength
and applied field. Note that the ground state crossings in Fig.
\ref{fig:fig3}(b) do not directly correspond to the lowest current
values in Fig. \ref{fig:fig3}(c), due to the energy contribution
of the Cooper-pairs distribution. Overall, one concludes that the
total current in the sample for given parameters is the
determining factor for the field-shift in the $j_{c}(H_{ext})$
characteristics.

\begin{figure}[t]
\onefigure[height=5cm]{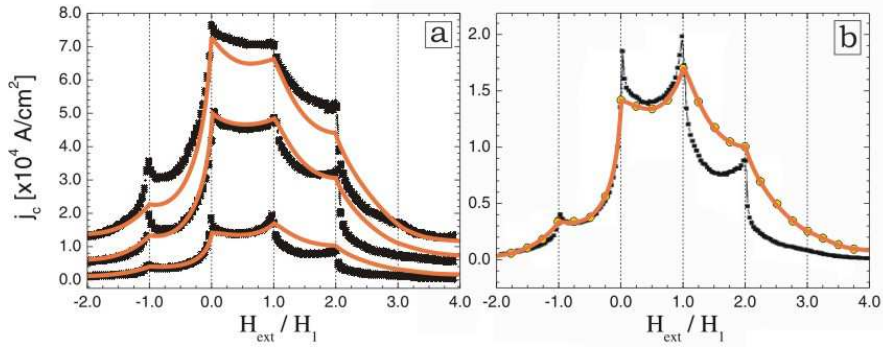} \caption{Critical current versus the
applied magnetic field in units of the first matching field
($H_{1}=57.5$G), for parameters taken from Ref. \cite{morgan} (a)
for temperatures $T/T_{c}=0.965$, $0.970$, and $0.980$ (top to
bottom); (b) enlargement of the $T/T_{c}=0.980$ data. Yellow dots
(and/or orange lines) denote the theoretical results. Experimental
data are from Ref. \cite{morgan}.} \label{fig:fig4}
\end{figure}

Interestingly enough, the phenomenon of the field-shift of the
critical current was experimentally observed in Ref.
\cite{morgan}, but not commented on. Namely, the authors were
concerned only about the field-polarity-dependent pinning, and
overlooked this issue. For comparison, we performed the
calculations for a SC film with a triangular MD-lattice on top,
where the parameters are taken from Ref. \cite{morgan} [array of
Ni dots ($R_{d}=120$nm, $D=110$nm, $W=0.6\mu$m) covered by a Nb
film ($d=95$nm), with $T_{c}\approx 8.60\pm 0.10$K]. Fig.
\ref{fig:fig4} shows the critical current in the sample as a
function of an applied field, at temperatures $T/T_{c} \approx
0.965$, $0.97$, and $0.98$ (orange lines) compared to experimental
data at $T=8.40$, $8.46$, and $T=8.52$K, respectively. As clearly
shown in Fig. \ref{fig:fig4}(b), the maximum in the critical
current shifts to the first matching field when approaching
$T_{c}$. In Fig. \ref{fig:fig3}(c), we demonstrated the
compensation of vortex-currents with $M$-dependent magnet-induced
currents. In the present case, for fixed magnetization $M$, the
vortex-currents profile is changed by temperature (which
effectively changes $\xi(T)$).

Our theoretical results, denoted by orange lines in Fig.
\ref{fig:fig4}, confirm the field-shift found experimentally. This
fit was obtained for $\xi(0)=36$nm and $\lambda(0)=90$nm values,
with $M=625$G, $20\%$ larger than the saturation magnetization of
bulk Ni ($510$G). From $H_{c2}$ measurements, the mean free path
was estimated in Ref. \cite{morgan} as $l \approx 5$nm (dirty
limit), which justifies the low $\xi(0)$ obtained theoretically.
Besides the agreement in the field shift in the critical current
dependence and the qualitative agreement, one should notice the
discrepancy in the critical current values. In our opinion, the
smearing of the matching peaks in our results is caused by the
large coherence length at temperatures close to $T_{c}$. $\xi(T)$
becomes comparable to the distance between the magnets, leading to
overlapping vortices. On one hand, this leads to enhanced
current-compensation effects, crucial for the field-shifted
$j_{c}(H_{ext})$ characteristics. However, this overlap results in
less effective matching effects (both integer and rational). Also,
one should note the partial `washing-out' of the magnetic
alignment when sweeping applied magnetic field in the experiment.
This leads to a less ordered vortex structure for non-matching
fields, effectively decreasing the critical current (i.e.
pronouncing peaks at matching fields). Despite these differences,
the experiment-theory agreement is apparent, demonstrating that
the predicted field-shift of the maximum of the critical current
is not sensitive neither to the geometry of the magnetic lattice
nor to the shape of the magnetic dots (Fig. \ref{fig:fig3} vs.
Fig. \ref{fig:fig4}).

In conclusion, we have shown that magnetic nanostructuring of the
SC film not only enhances its critical field in a controlled
fashion, but also can enhance the maximal current the sample can
sustain. Moreover, the magnetic field for which the maximal
critical current is achieved can also be engineered. Our findings
are in excellent agreement with existing experiments
\cite{morgan,lange}. We interpret experimental results and our
predictions through the interaction of the external flux lines
with the magnet-induced currents (and vortex-antivortex pairs) in
the SC. These interactions may lead to novel vortex-antivortex
lattices with positive net vorticity.

\section{Acknowledgements}

The authors acknowledge D. Vodolazov, V. Moshchalkov and M. Van
Bael for valuable discussions. This work was supported by the
Flemish Science Foundation (FWO-Vl), the Belgian Science Policy,
and the University of Antwerp (GOA).

\end{document}